\documentclass[usenatbib,usegraphicx]{mn2e}
\usepackage{amsmath}
\usepackage[]{hyperref}
\usepackage{graphicx}
\usepackage{color}
\usepackage{units}
\usepackage{ulem}
\DeclareGraphicsExtensions{.eps,.pdf,.png,.jpg}
\hypersetup{
    colorlinks=true,
    linkcolor=black,
    citecolor=black,
    filecolor=black,
    urlcolor=black,
}

\begin{document}

%\title[TBD]{Planetary atmospheres are subject to evaporation from their host galaxy's supermassive black hole.}
\title[Black Holes vs Planets]{Evaporation of planetary atmospheres due to XUV illumination by quasars}

\author[J. C. Forbes and A. Loeb]{John C. Forbes,$^1$\textsuperscript{\thanks{E-mail: john.forbes@cfa.harvard.edu}}  Abraham Loeb$^1$  \\
$^1$Institute for Theory and Computation, Harvard University, Cambridge MA 02138, USA
}
%\author{John C. Forbes, Avi Loeb}

\maketitle

\begin{abstract}
Planetary atmospheres are subject to mass loss through a variety of mechanisms including irradiation by XUV photons from their host star. Here we explore the consequences of XUV irradiation by supermassive black holes as they grow by the accretion of gas in galactic nuclei. Based on the mass distribution of stars in galactic bulges and disks and the luminosity history of individual black holes, we estimate the probability distribution function of XUV fluences as a function of galaxy halo mass, redshift, and stellar component. We find that about 50\% of all planets in the universe may lose the equivalent of a Martian atmosphere, 10\% may lose an Earth's atmosphere, and 0.2\% may lose the mass of Earth's oceans. The fractions are appreciably higher in the spheroidal components of galaxies, and depend strongly on galaxy mass, but only weakly on redshift.
\end{abstract}

\section{Introduction}

The recent discoveries of the TRAPPIST1 planetary system \citep{gillon_seven_2017}, LHS 1140b \citep{dittmann_temperate_2017}, and Proxima Centauri b \citep{anglada-escude_terrestrial_2016}, each with planets in the putative habitable zone of their low-mass host stars, has generated intense interest in the susceptibility of their atmospheres to erosion and more generally the viability of life on their surfaces \citep{ribas_habitability_2016, garraffo_space_2016, airapetian_how_2017, dong_is_2017, ranjan_surface_2017}. Despite their low masses, M-dwarfs are capable of generating substantial high-energy radiation through frequent flares \citep[e.g.][]{barnes_habitability_2016, meadows_habitability_2016, davenport_most_2016, vida_frequent_2017, ribas_full_2017, miles_hazmat_2017}. The large flux of high-energy radiation to which the close-in planets are subjected is capable of ionizing their upper atmospheres and driving hydrodynamic winds, the same mechanism responsible for occasionally dramatic mass loss among hot Jupiters \citep{murray-clay_atmospheric_2009}.

A given planet may be subject to sources of high energy radiation besides its host star. There are many such sources in the universe, but we will argue that the most important, aside from the host star of the planet, is radiation from the growth of supermassive black holes in galactic nuclei. In their most extreme phase of growth, when the black hole is potentially visible as a quasar, we will argue that the radiation from the black hole is sufficient to substantially affect planets on scales comparable to the entire galaxy.

In this paper we will apply results from the study of stellar XUV radiation on planetary atmospheres to the problem of quasars illuminating the same atmospheres. This will result in a straightforward relationship between the planet's distance from the center of the galaxy, the mass of the supermassive black hole at the galaxy's center, and the mass lost by the planet's atmosphere. Based on this relationship, we can estimate the probability distribution function of mass loss, i.e. the probability a random planet in the universe will be subject to a given amount of mass loss, as a function of redshift, mass, and the stellar component in which the planet's host star resides. We will estimate the effect introduced by the fact that many stars in the universe have non-circular orbits, and hence will occasionally pass much closer to the central SMBH. Finally, we discuss the implications for habitability - on Earth, and in galaxies in general.

\section{The relationship between XUV flux from black holes and mass loss from planetary atmospheres}

Planetary atmospheres subjected to the high-energy radiation of Active Galactic Nuclei (AGN) may experience a number of effects harmful to life on the planet. The primary effect we consider here is the loss of substantial mass from the atmosphere due to heating by photons near or exceeding the ionization potential of hydrogen.

In principle the mass loss from planetary atmospheres can be quite complicated and difficult to predict. In the solar system, the Earth loses mass via Jeans mass loss, namely the loss of individual particles on the high-velocity tail of the Maxwell-Boltzmann distribution, while Mars has lost much of its atmosphere to the solar wind via sputtering as a result of a minimal protective magnetic field and low surface gravity \citep{jakosky_mars_2017}. In addition to mass lost one particle at a time by these methods, planets subject to heavy irradiation from high-energy photons can experience hydrodynamic winds, wherein the upper atmosphere is heated and flows outward in the fluid regime \citep[e.g.][]{lammer_atmospheric_2003, erkaev_roche_2007, murray-clay_atmospheric_2009, lopez_how_2012, owen_planetary_2012, owen_uv_2016, chen_evolutionary_2016, zahnle_cosmic_2017}.

As is standard in the literature, we parameterize the mass loss by the incident flux of XUV radiation (meaning radiation with energies encompassing extreme ultraviolet radiation starting at 10 eV up to and including soft X-rays),
\begin{equation}
\dot{M} = \epsilon \frac{F_\mathrm{XUV} \pi R_p^2}{G M_p/R_p}
\label{eq:mdot}
\end{equation}
Here $F_\mathrm{XUV}$ is the XUV flux, and $R_p$ and $M_p$ are the plaent's radius and mass respectively. Physically, this equation assumes that some fraction $\epsilon$ of the incident power available to heat the atmosphere, $\pi R_p^2 F_\mathrm{XUV}$, is used in driving the atmosphere out of the planet's gravitational potential. Crucially $\dot{M}$ depends on the planet's properties (excluding composition and other higher-order complications) only via the bulk density $\rho_p \propto M_p/R_p^3$, which does not vary immensely, particularly for rocky planets of the most interest for habitability.

The value of $\epsilon$ can be estimated theoretically \citep{murray-clay_atmospheric_2009, owen_uv_2016} to be of order $10\%$ over a wide range of fluxes \citep{bolmont_water_2017}. Below a critical value of $\sim 0.1\ \mathrm{erg}\ \mathrm{s}^{-1}\mathrm{cm}^{-2}$, the flux is insufficient to drive a hydrodynamic wind and $\epsilon$ falls precipitously. Above about $1000\ \mathrm{erg}\ \mathrm{s}^{-1}\mathrm{cm}^{-2}$, $\epsilon$ begins to fall slowly as the outflow solution begins to lose substantial energy to radiative cooling. In this regime $\dot{M}\propto F_\mathrm{XUV}^{1/2}$ \citep{murray-clay_atmospheric_2009}. The Appendix discusses an example where the outflow is subject to high radiative losses, namely planets in the habitable zone around host objects with high XUV luminosities. In the regime where $\epsilon\approx 10\% \approx \mathrm{const.}$, one only needs a timescale over which a given XUV flux acts to estimate the total mass lost by hydrodynamic winds.

The total mass lost due to hydrodynamical winds in the regime where $\epsilon\approx\mathrm{const}$ is simply the time integral of equation \eqref{eq:mdot},
\begin{equation}
\label{eq:mlost}
M_\mathrm{lost} \approx 3\epsilon (4\pi G\rho_P)^{-1}  \int F_X dt
\end{equation}
where the integral is evaluated from the time a planet develops a steady atmosphere to the epoch under consideration. We have assumed that the planet's properties are roughly constant in time. This may not be true in the early stages of planetary evolution, as even terrestrial planets may retain substantial gas accreted directly from the protoplanetary disk. However, we expect that for planets that may host life, the properties of the planet itself should change relatively little over long time periods.

Equation \eqref{eq:mlost} implies that the critical remaining unknown quantity is the fluence, $\Psi_X \equiv \int F_X dt$, integrated over the relevant timescale. In order to proceed, one path is to specify a source luminosity $L_X$ and lifetime $t_0$, which translates into a fluence given a distance $r$, i.e. $\Psi_X = t_0 L_X / (4\pi r^2)$. In the case of a supermassive black hole, we take $L_X = \eta_X \eta_\mathrm{Edd} L_\mathrm{Edd}$, where $\eta_X$ is the fraction of the spectrum of sufficiently high energy to aid in driving hydrodynamic winds, $\eta_\mathrm{Edd}$ is the Eddington ratio, namely the ratio of the bolometric luminosity to the Eddington luminosity, and $L_\mathrm{Edd} \approx 1.3 \times 10^{38} (M_\mathrm{BH}/M_\odot)\ \mathrm{erg}\ \mathrm{s}^{-1}$ is the Eddington luminosity itself for a black hole of mass $M_\mathrm{BH}$.

The value of $t_0$ has been extensively studied over many decades as the quasar duty cycle or lifetime. It can be estimated by matching the present-day mass function of supermassive black holes to the luminosity function of quasars at higher redshift \citep{richstone_supermassive_1998, haehnelt_high-redshift_1998, salucci_mass_1999, wyithe_physical_2002, aird_x-rays_2017} the clustering properties of quasars \citep{martini_quasar_2001}, or via hydrodynamic simulations \citep{hopkins_physical_2005}. These models are generally consistent with a picture in which the black hole radiates at or near the Eddington luminosity during the quasar phase for a time of order the Salpeter time, i.e. the mass-independent time it takes to double the black hole's mass during Eddington-limited accretion with radiative efficiency $\epsilon_\mathrm{BH}\sim 10\%$, roughly $t_0 = 40\ \mathrm{Myr}$. In this case, $\Psi_X = t_0 \eta_X L_\mathrm{edd} / (4\pi r^2)$.

At the same time, we note that the current mass of the black hole may be written directly in terms of the fluence as follows.
\begin{equation}
\label{eq:MBH}
M_{BH} =\int \frac{(1-\epsilon_\mathrm{BH}) L_{BH} }{ \epsilon_\mathrm{BH} c^2} dt =\frac{4\pi r^2}{c^2} \int \frac{ (1-\epsilon_\mathrm{BH}) F_X}{\epsilon_\mathrm{BH} \eta_X} dt
\end{equation}
The remaining fraction, $1-\epsilon_\mathrm{BH}$, of this energy is gained by the black hole thereby increasing its mass. Canonically $\epsilon_\mathrm{BH} \sim 0.1$. If we assume that both $\epsilon_\mathrm{BH}$ and $\eta_X$ are constant over the lifetime of the black hole, we can rearrange the previous equation to obtain
\begin{equation}
\Psi_X \approx \frac{\epsilon_\mathrm{BH}}{1-\epsilon_\mathrm{BH}} \frac{\eta_X M_\mathrm{BH} c^2 }{  4\pi r^2 }
\end{equation}
So long as $\eta_X$ and $\epsilon_\mathrm{BH}$ are roughly constant, this expression will hold regardless of the growth history of the black hole. This is a nice feature when compared to our previous estimate of $\Psi_X$, which requires explicitly assuming that the black hole accretes at the Eddington-limited rate for a fixed time $t_0$. However, the two estimates are quite closely related, since some methods for estimating $t_0$ rely on equation \eqref{eq:MBH}. Moreover, the two estimates depend on $r$, $M_\mathrm{BH}$, and $\eta_X$ in exactly the same way, so we can find a $t_0$ for which the two are identical for any given value of $\epsilon_\mathrm{BH}$. For the canonical value of $\epsilon_\mathrm{BH}=0.1$, the fluences are equal when $t_0=50$ Myr, remarkably close to our fiducial choice above. 

Because these estimates are so congruous, for convenience and definitiveness we will adopt $t_0=40\ \mathrm{Myr}$. This will allow us to make a more sophisticated estimate of $\eta_X$, since we will know in what state the black hole is accreting. This will also allow us to easily check which fluences correspond to fluxes above or below the critical values for $\epsilon$ discussed above. A corollary is that fluxes, fluences, and total mass lost are directly proportional to each other over several orders of magnitude in each.

In Table \ref{tab:sources}, we list various masses whose loss may be important for a planet's habitability, and the corresponding high-energy fluences and fluxes that would be required to remove that much mass assuming $\Psi_X = F_X t_0$. One can also go in the other direction, starting with a known flux, such as the high-energy flux at Proxima Centauri \citep{ribas_full_2017}, or the current high energy flux at the earth from the sun \citep{ribas_evolution_2005}, or from Sgr A* at Eddington luminosity. To translate between fluxes and fluences, we usually assume our fiducial value of $t_0 = 40\ \mathrm{Myr}$, but for cases in the table where the high-energy flux originates from a star, we adopt a substantially longer timescale of $5\ \mathrm{Gyr}$ to reflect the fact that the planets in question have experienced that flux for a much longer period of time. For the purposes of these simple estimates, we neglect the fact that younger stars have substantially higher fractions of their luminosity emitted in the XUV \citep{ribas_evolution_2005} -- for the sun this increases the XUV fluence by about an order of magnitude \citep{zahnle_cosmic_2017}. Values for the current XUV fluxes at the Earth from the Sun, and at Proxima b from Proxima Centauri are taken from \citet{ribas_evolution_2005} and \citet{ribas_full_2017} respectively. %We also include the minimum and maximum fluxes expected to be required for efficient hydrodynamical mass loss \citep{owen_uv_2016, bolmont_water_2017}

\begin{table*}
\caption{Examples of fluxes, fluences, mass loss rates, and accumulated mass lost. In the top half of the Table, we use a timescale appropriate for the growth of supermassive black holes (40 Myr), while for the bottom half we use a timescale appropriate for planets around older stars (5 Gyr). Bold entries indicate the starting point from which the other three entries are derived for a given row.}
\begin{tabular}{lccccc}
  & $\log_{10}F_\mathrm{XUV}\ \left(\frac{\mathrm{erg}}{\mathrm{s}\ \mathrm{cm}^2} \right)$ & $\log_{10} \Psi_X\ \left(\frac{\mathrm{erg}}{\mathrm{cm}^2}\right)$ & $\log_{10} \dot{M}_\mathrm{loss}$  $\left(\frac{\mathrm{g}}{\mathrm{s}}\right)$ & $\log_{10}M_\mathrm{lost}$ (g)  \\
  \hline
  \hline
Earth's Oceans & 3.7 &  18.8 & 9.0 & {\bf 24.15} \\
Earth's Atmosphere & 1.3 & 16.4 & 6.6 & {\bf 21.7} \\
Mars's Atmosphere & -1.0 & 14.1 & 4.3 & {\bf 19.4} \\
Flux at Earth from Sgr A* if $\eta_X = 0.1$ & {\bf -2.18} & 12.9 & - & - \\
\hline \\
Current XUV Flux at Proxima b & {\bf 2.47} & 19.7 & 7.8 & 25.0 \\
Current XUV Flux at Earth & {\bf  0.67} & 17.9 & 6.0 & 23.2 \\
Current Earth Mass Loss Rate & -1.8 & 15.4 & {\bf 3.5} &  20.7 \\
\hline \\
\end{tabular}
\label{tab:sources}
\end{table*}

We shall use the first three entries in the Table as references in the figures throughout the rest of the paper. In addition to being a useful reference, this Table also immediately raises an important issue. The current mass loss rate from the Earth's atmosphere is a factor of 300 lower than predicted by Equation \eqref{eq:mdot}, despite the fact that the current XUV flux at the Earth is well within the range for which \citet{bolmont_water_2017} predict that $\epsilon\sim 0.1$. This is understood to be a feature of the current structure of Earth's atmosphere, wherein hydrogen and helium only escape by diffusion. With warmer, still formally habitable, surface temperatures, substantially more water vapor could make it to the upper atmosphere where it would be subject to photodissociation and the loss of hydrogen \citep{kasting_stratospheric_2015}. This discrepancy emphasizes the importance of other factors in the structure of the atmosphere which we do not consider. It is therefore prudent to expect that our results will only hold for the subset of planetary atmospheres to which Equation \eqref{eq:mdot} applies.

\section{The distribution of xuv fluences from AGN}

In the previous section we showed that the mass lost from a given planet's atmosphere satisfies $M_\mathrm{lost} \propto \Psi_X = t_0 \eta_X \eta_\mathrm{Edd} L_\mathrm{Edd} / (4\pi r^2)$, where $t_0$, $\eta_X$, and $\eta_\mathrm{Edd}$ can each be estimated. This relation can be re-arranged to obtain the three-dimensional distance $r$ between the planet and the black hole at which the planet is subject to a given fluence for a particular choice of $E_X \equiv t_0 \eta_X \eta_\mathrm{Edd} L_\mathrm{Edd}$. Scaling to relevant physical values, 
 \begin{equation}
	 \label{eq:rcrit}
r = 1.8\ \mathrm{kpc} \left(\frac{M_\mathrm{BH}}{10^9 M_\odot} \right)^{1/2} \left( \frac{\Psi_X}{10^{17} \mathrm{erg}\ \mathrm{cm}^{-2}} \right)^{-1/2}
\end{equation}
Clearly this set of values suggests that this effect can be important, at least in the cores of galaxies. 
\begin{figure}
\centering
\includegraphics[width=3.8in]{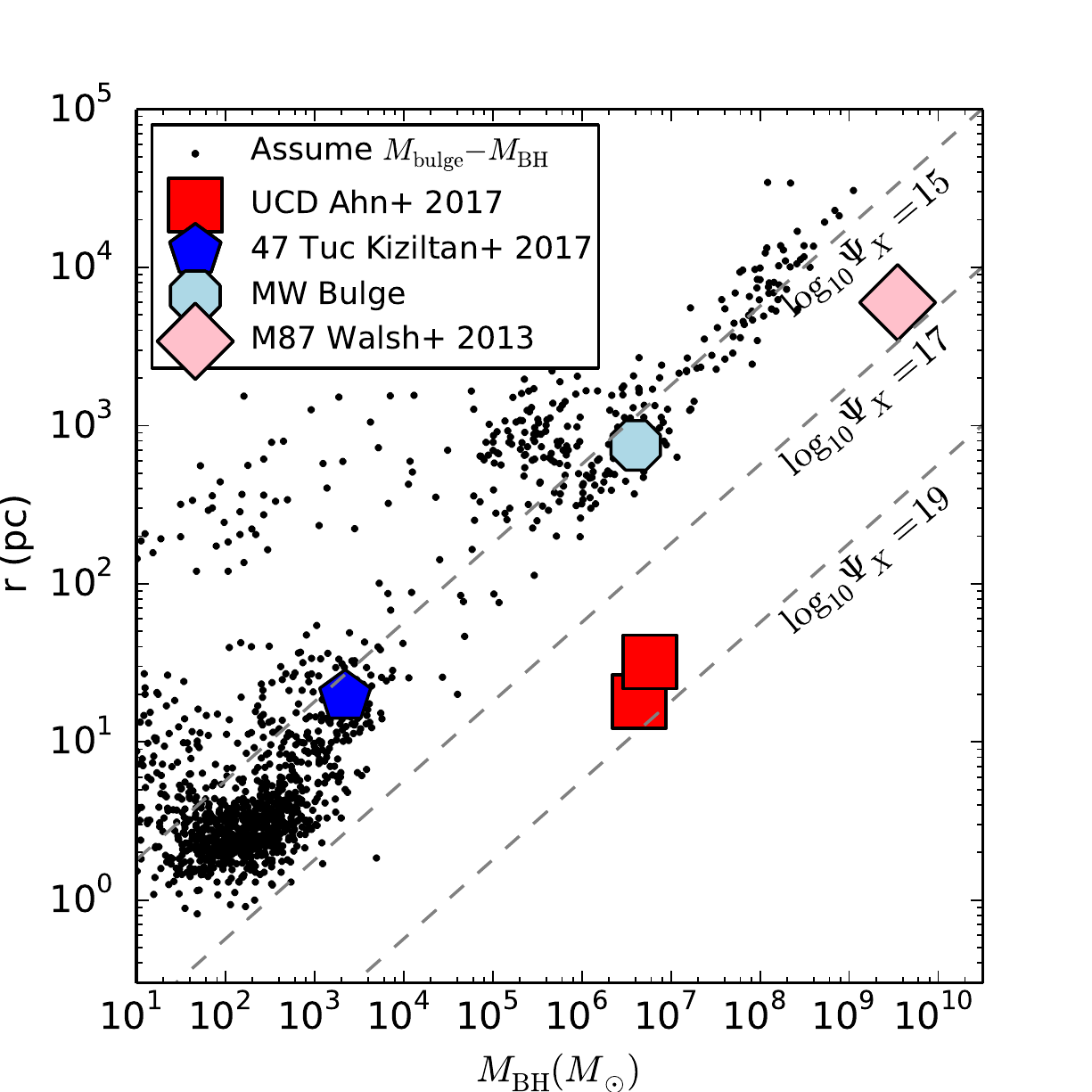}
\caption{Black hole mass vs. stellar half-light radius for spheroidal stellar systems. Black points are drawn from the SAGES database of half-light radius vs V-band luminosity, and translated into a black hole mass assuming the correlation from \citet{haring_black_2004}, and so should be interpreted with caution especially below $M_\mathrm{BH} \sim 10^7 M_\odot$. Colored points show individual systems where the black hole mass is known, and the diagonal dashed lines show the radii at which different XUV fluences would be experienced.}
\label{fig:everythingPlot}
\end{figure}

As a first estimate of how relevant this effect is, we adapt a plot from \citet{brodie_relationships_2011}, namely masses of spherical stellar systems plotted against their effective radii. Lines of constant $\Psi_X$ are superimposed, suggesting which systems will contain large fractions of planets strongly effected by XUV radiation from their supermassive black hole. Figure \ref{fig:everythingPlot} shows the result, where a few systems of known black hole mass and half light radii are highlighted in color. The smaller background points show a large compilation of data spanning globular clusters to elliptical galaxies\footnote{\url{http://sages.ucolick.org/spectral_database.html} retrieved May 10, 2017} \citep[including data from][and many others]{barmby_structural_2007, chilingarian_population_2009, harris_new_2010, brodie_relationships_2011, strader_wide-field_2011, misgeld_families_2011, mcconnachie_observed_2012, usher_sluggs_2013, jennings_sluggs_2014}, where their known V-band luminosities have been translated into black hole masses by assuming that the correlation between spheroid luminosity and black hole mass \citep[here we use][]{haring_black_2004} established at high masses extends to arbitrarily small masses. This is a highly speculative assumption \citep[see e.g.][for steeper versions of this relation]{kormendy_coevolution_2013, graham_black_2015, pihajoki_geometric_2017}.

We highlight a few systems in Figure \ref{fig:everythingPlot} that do not require translation through this empirical correlation because their black hole masses are estimated based on direct modeling of the system in question. For the Milky Way, we assume a half mass radius of $0.75\ \mathrm{kpc}$ for the bulge \citep[e.g.][]{li_modelling_2016}, and a black hole mass of $4 \times 10^{6} M_\odot$ \citep[e.g.][]{boehle_improved_2016}. For M87, we take the black hole mass from \citet{walsh_m87_2013} and the effective radius from \citet{agnello_dynamical_2014}. We also plot the recently-reported intermediate mass black hole from in 47 Tuc \citet{kiziltan_intermediate-mass_2017} and two black holes in the ultra-compact dwarf galaxies reported in \citet{ahn_detection_2017}. 

Based purely on Figure \ref{fig:everythingPlot}, focusing largely on the regime $M_\mathrm{BH}\ga 10^6 M_\odot$ where the relationship between luminosity and black hole mass is reasonably well-callibrated, we can already see a few important results. The half-mass radii of these stellar systems are generally large enough that many of the planets they presumably contain will be subject to only moderate quantities of XUV fluence -- enough to damage the present-day atmosphere of Mars, but not much more. There are several interesting outliers, namely M87, and the ultra-compact dwarfs. These are each peculiar systems in their own way, respectively one of the most massive galaxies in the local universe, and galaxies with remarkably small radii. This suggests that regardless of the fate of the typical planet in the universe, there will be some unlucky enough to reside in unusual galaxies such as these. Globular clusters also stray towards the $\log_{10} \Psi_X = 17$ line, suggesting that globular clusters may be less hospitable to life than the typical star. Of course it is not clear that any of them actually contain the intermediate mass black holes suggested by the extrapolation, but interestingly the reported result of \citet{kiziltan_intermediate-mass_2017} is very much in line with the extrapolation.

Figure \ref{fig:everythingPlot} is a simple first look at how planets in nearby stellar systems might fare under the influence of XUV radiation from their black holes. To take it a step further, we will model the full stellar distributions (rather than simply looking at half-light radii), in order to compute the probability density that a random planet at a particular epoch in the universe will be subject to a fluence $\Psi_X$. The first step is to compute the distribution $dP/d\log_{10}\Psi_X$ for a single stellar component. With these distributions, we can then construct the full distribution of expected fluences for a random planet in the universe (as opposed to a random planet in a given stellar system) by summing up the individual distributions with appropriate weights.

The fluence distribution for a given stellar system is straightforward to compute. Given that there is a relatively simple relationship between $r$ and $\Psi_X$, namely Equation \eqref{eq:rcrit}, we can change variables so that
\begin{equation}
\label{eq:dmdpsi}
\frac{dM_*}{d\Psi_X} = \frac{dM_*}{dr} \left| \frac{ dr}{d\Psi_X} \right|
\end{equation}
where $M_*$ is the cumulative mass in stars, either exposed to a given $\Psi_X$ or within radius $r$. The radial mass distribution $dM_*/dr$ has a number of well-known forms, and $dr/d\Psi_X$ follows immediately from equation \eqref{eq:rcrit}. We should note that in every plot in this paper, we are actually showing either $dM_*/d\log_{10}\Psi_X = \ln(10)\Psi_X dM_*/d\Psi_X$, or the probability density per $\log_{10} \Psi_X$, namely $dP/d\log_{10} \Psi_X = (1/M_\mathrm{tot}) dM_*/d\log_{10}\Psi_X$, since the $x$-axis of each plot will be logarithmically scaled. Here $M_\mathrm{tot}$ is the total stellar mass in a given system.

\begin{figure*}
\centering
\includegraphics[width=6in]{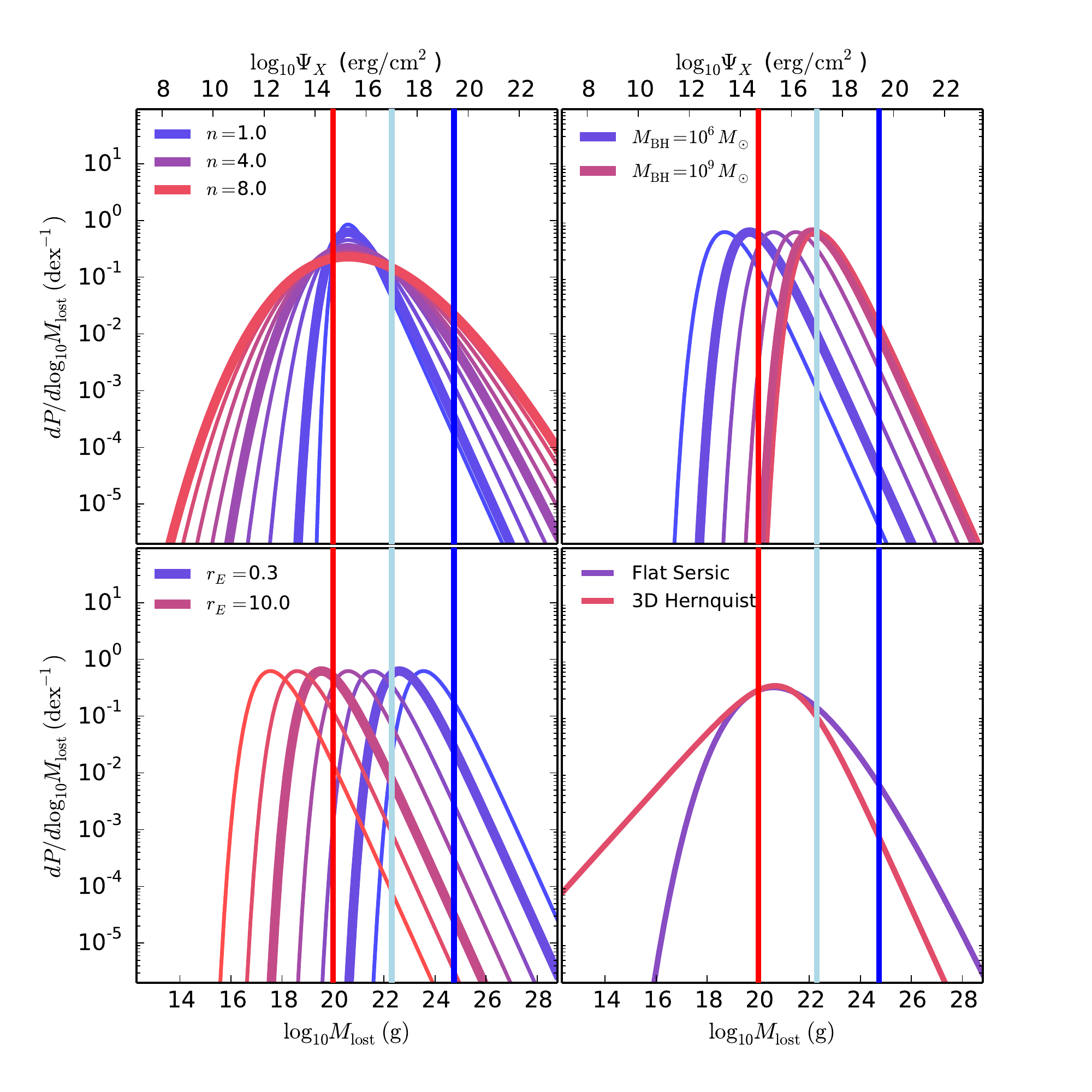}
\caption{The atmospheric mass loss distribution for different radial profiles of stars in their host galaxies. The vertical lines show reference values of the mass lost from the planet's atmosphere, $M_\mathrm{lost}$ or equivalently the XUV fluence $\Psi_X$, corresponding to the masses of the Martian atmosphere, Earth's atmosphere, and the mass in Earth's oceans (red, light blue, and dark blue respectively). Each panel shows the fluence probability distribution function (PDF) varying one parameter at a time, and keeping the others fixed at $M_\mathrm{BH} = 10^7 M_\odot$, a Sersic index $n=1$, or an effective radius $r_E=3\ \mathrm{kpc}$. In the bottom right panel, we compare a razor-thin $n=4$ distribution to a more realistic Hernquist distribution.}
\label{fig:demoplot}
\end{figure*}

For a thin axisymmetric disk with surface density $\Sigma$, we have $dM_*/dr = 2\pi r \Sigma$, while for a spherical distribution with density $\rho$, $dM_*/dr = 4\pi r^2 \rho$. A standard choice for $\Sigma$ is the Sersic density profile \citep{sersic_influence_1963}, 
\begin{equation}
\Sigma = \Sigma_0 \exp\left( -b(r/r_E)^{1/n} \right)
\end{equation}
where $\Sigma_0$ is a normalization constant, $r$ is the distance from the center of the galaxy, $r_E$ is the radius containing half the mass, $n$ is the Sersic index, and $b$ is a function of $n$ set such that $r_E$ does in fact contain half the mass. Plugging this profile into equation \eqref{eq:dmdpsi} and adding the appropriate factors to change variables to log-scale, we find
\begin{equation}
\frac{dP}{d\log_{10} \Psi_X} = \frac{\ln(10)}{\Gamma(2n)} \frac{E_X b^{2n}}{8\pi  r_E^2 n \Psi_X} \exp\left(-b \left(\frac{E_X}{4\pi  r_E^2 \Psi_X}\right)^{1/2n}  \right)
\label{eq:fluencedist}
\end{equation}
Here $\Gamma(x) = \int_0^\infty t^{x-1}e^-t dt$ is the gamma function. This equation is quite straightforward to evaluate, and essentially reduces to a power law with an exponential cutoff in the quantity $(E_X/r_E^2 \Psi_X)^{(1/2n)}$. This immediately says that changes in $E_X$ and $r_E^2$ affect the distribution in the same way but with opposite sign -- increasing $E_X$ or decreasing $r_E^2$ shift the distribution to higher $\Psi_X$. Increasing $n$, meanwhile, increases the width of the distribution. Each of these effects is shown in turn in Figure \ref{fig:demoplot}.

Figure \ref{fig:demoplot} also shows the difference between the infinitely thin Sersic profile with $n=4$, and the three-dimensional \citet{hernquist_analytical_1990} profile, which when viewed in projection yields an excellent approximation to the $n=4$ Sersic profile. Since the two profiles are nearly the same in projection, clearly each particle in the 3D distribution will be strictly farther away from the center than the corresponding particle in the flattened distribution. It is therefore unsurprising that the flat distribution extends to higher $\Psi_X$, while the 3D distribution contains contributions at much lower $\Psi_X$. Following the same procedure as for the Sersic profile, but with $dM_*/dr = 4\pi r^2 \rho$, and $\rho=M_\mathrm{tot} a / (2\pi r (r+a)^3)$, we find that the fluence distribution is
\begin{equation}
\label{eq:fluencedistHernquist}
\frac{dP}{d\log_{10}\Psi_X} = \frac{\ln(10) E_X / (4\pi \Psi_X a^2)}{ (\sqrt{E_X/(4\pi a^2 \Psi_X)} + 1)^3}
\end{equation}
Here $a$ is the characteristic scale of the potential-density pair, which is related to the effective radius (i.e. the projected half-mass radius) via $r_E \approx 1.8153 a$ \citep{hernquist_analytical_1990}. Just as in equation \eqref{eq:fluencedist}, the dimensionless quantity determining the shape of the distribution is $E_X/(4\pi r_E^2 \Psi_X)$, which is the ratio of the fluence in question $\Psi_X$ to the fluence at the half-mass radius $r_E$.

While these distributions show the fluence expected for a single component of a single galaxy, a random planet in the universe will be located in a multi-component galaxy with a distribution of values of $r_E$, $n$ and $E_X$. To estimate the fluence distribution for all planets, we proceed via the Monte Carlo method in the spirit of \citet{guillochon_fastest_2015}, namely drawing values of each of these quantities from an appropriate distribution and summing the results.

Explicitly, we proceed using the following notation: $M_h$ is the mass of the dark matter halo, $M_*$ is the stellar mass, $M_b$ is the bulge mass, $M_d$ is the disk mass, $SF$ is either true or false, true meaning that the galaxy is star-forming and false meaning that the galaxy is quenched, the bulge to total ratio is $BT=M_b/(M_b+M_d)$, and $r_{E,d}$ and $r_{E,b}$ are the effective radii of the disk and bulge respectively. Our calculation proceeds as follows:
\begin{enumerate}
\item Set up a grid of halo masses, equally spaced in $\log_{10} M_h$ with spacing $\Delta\log_{10} M_h$. For each $M_h$...
\item Record the density of halos per dex in halo mass, $\phi_{10}(M_h)$
\item Draw from the conditional probability distribution $p(M_* | M_h, z)$
\item Draw from $P(SF | M_*, z)$
\item Draw from $p(BT | M_*, SF)$, implying $M_d$ and $M_b$.
\item Draw from $p(M_\mathrm{BH} | M_b)$, implying $E_X$.
\item Draw from $p(r_{E,d} | M_d, SF, z )$
\item Draw from $p(r_{E,b} | M_b, SF, z )$
\item Use Equations \eqref{eq:fluencedist} and \eqref{eq:fluencedistHernquist} to compute $dP/d\log_{10}\Psi$ for both the disk and bulge component respectively.
\end{enumerate}
Each of these distributions is then summed up with the appropriate weight to yield the density of stellar mass in the universe subject to a given fluence, $d \rho_*/d \log_{10} \Psi$, which has units of solar masses per cubic Mpc per dex in fluence,
\begin{eqnarray}
	\label{eq:sum}
\frac{ d \rho_*}{d\log_{10} \Psi} &\approx & \sum_i \Bigg[ \phi_{10}(M_{h,i}) \Delta\log_{10} M_h \times  \\
& & \ \ \left(M_d \frac{dP}{d\log_{10}\Psi} \Big|_d  + M_b \frac{dP}{d\log_{10}\Psi}\Big|_b \right) \Bigg] \nonumber
\end{eqnarray}
This quantity is proportional to the number of planets per volume in the universe per dex in fluence, assuming that planet formation has no dependence on additional parameters. The validity of this assumption and the consequences of relaxing it will be discussed in section \ref{sec:discussion}.

In order to proceed, we need to specify the various probability density functions in the list above. For the most part these will be normal distributions in the logarithm of the variables, reflecting the fact that observationally many of these relations are consistent with having a fixed lognormal scatter. 

First, we have the halo mass function $\phi_{10}(M_h)$, where the subscript is a reminder that this quantity is $dN/d\log_{10} M_h$, not $dN/dM_h$. For these we use the online tool HMFcalc \citep{murray_hmfcalc:_2013} with the default settings, namely a Planck-SMT cosmology \citep{planck_collaboration_planck_2016} cosmology and a \citet{sheth_ellipsoidal_2001} fitting function, but we expand the tabulated range of masses to $10^8$ to $10^{17} M_\odot/h$, decrease the tabulated mass intervals, and obtain values at redshifts between $z=0$ and $z=10$ in steps of $\Delta z =1$. The value of the mass function here simply serves as a weight in the sum given by equation \eqref{eq:sum}.

For the distribution of stellar masses given a halo mass, $p(M_* | M_h)$, we employ the redshift-dependent relation from \citet{moster_galactic_2013}, with a log-normal scatter of $0.15 \mathrm{dex}$ independent of $M_h$ or $z$. The scatter in this relation has been constrained with a variety of techniques at $z=0$ \citep{reddick_connection_2013, zu_mapping_2015}, but its redshift evolution is not well-constrained. The relationship between scatter and the low-mass slope of the stellar mass halo mass relation is also degenerate - that is, the data are consistent with a family of solutions where the slope and scatter are increased simultaneously \citep{garrison-kimmel_organized_2017}. However, as we will see, the low-mass end of the relation is not incredibly important to our final result.

Next, we draw a random number to determine if the galaxy is star-forming or quenched. To do so we employ the stellar mass function fits provided by \citet{leja_reconciling_2015}. In particular the stellar mass function is split into star-forming and quenched galaxies, so the probability that a galaxy is star-forming or quenched as a function of stellar mass is just $P_\mathrm{SF} = \phi_{SF}(M_*)/(\phi_{SF}(M_*) + \phi_\mathrm{quenched}(M_*))$.

Once the star-forming status of the galaxy is known, we can draw from the observed relationship between bulge mass and stellar mass from \citet{lang_bulge_2014}, which shows little dependence on redshift. For star-forming galaxies,
\begin{equation}
	BT \sim \begin{cases} \mathcal{N}_{0.01-0.99}\left(  0.2 + 0.2 \log_{10} \left(\frac{M_*}{10^{10} M_\odot}\right) , 0.33 \right) \ \ \ \mathrm{if} \ \mathrm{SF} \\
	\mathcal{N}_{0.01-0.99}\left( 0.55 + 0.1 \log_{10} \left(\frac{M_*}{10^{10} M_\odot}\right) , 0.33 \right) \ \mathrm{if}\ \mathrm{not}\ \mathrm{SF},
	\end{cases}
\end{equation}
meaning that $BT$ is drawn from a clipped normal distribution, which is a normal distribution with the mean given by the first argument and standard deviation given by the second. The allowed range is given by the subscript, and if a value is drawn outside the allowed range, it is replaced with the appropriate bound, e.g. $BT = 0.01 $ if the draw from the normal distribution is below $0.01$.

With the mass partitioned, we rely on the observed correlation between bulge mass and black hole mass. In particular we adopt the relation reported in \citet{mcconnell_revisiting_2013}, namely
\begin{equation}
	\log_{10} M_\mathrm{BH} \sim \mathcal{N}( 8.46 + 1.05 \log_{10} \left(\frac{M_b}{10^{11} M_\odot}\right), 0.34)
\end{equation}
where $\mathcal{N}$ with no subscript denotes a normal distribution, meaning that $M_\mathrm{BH}$ itself is drawn from a log-normal distribution. Note that the scatter in the relation is quite uncertain, with values quoted between 0.2 and 0.6 depending on the statistical method, and with some indication that the scatter is larger at smaller masses. With the mass of the black hole, we can now estimate $E_X = t_0 \eta_X \eta_\mathrm{Edd}  L_\mathrm{Edd}$. As discussed in the previous section, we take $t_0 = 40\ \mathrm{Myr}$, $\eta_\mathrm{Edd}=1$, and $L_\mathrm{Edd} = 1.3 \times 10^{38} (M_\mathrm{BH}/M_\odot)\ \mathrm{erg}\ \mathrm{s}^{-1}$. For $\eta_X$, rather than assuming a constant value, we adopt a model for the spectrum of AGN from \citet{thomas_physically-based_2016}, assuming in their model that the proportion of flux in a non-thermal component is $p_{NT}=0.1$, the photon power-las index of the non-thermal emission is $\Gamma=1.8$, and the coronal radius in units of gravitational radii is $\log_{10}r_\mathrm{cor}=1.5$.

The final ingredients entering our model for the cosmological fluence distribution are the effective radii of each component. Once again we rely on the results of \citet{lang_bulge_2014}
\begin{equation}
	\label{eq:rEb}
	\log_{10} r_{E,b} \sim \begin{cases} \mathcal{N} \left(  -0.097 + 0.097 \log_{10} \left(\frac{M_*}{10^{10} M_\odot}\right) , 0.38 \right) \ \ \mathrm{if} \ \mathrm{SF} \\
	\mathcal{N} \left( -0.35 + 0.25 \log_{10} \left(\frac{M_*}{10^{10} M_\odot}\right) , 0.38 \right)\ \mathrm{not}\ \mathrm{if}\ \mathrm{SF}
	\end{cases}
\end{equation}
for the spheroidal component, or
\begin{eqnarray}
\label{eq:rEd}
	\log_{10} r_{E,d} \sim \begin{cases} \mathcal{N}\left(  0.48 + 0.125 \log_{10} \left(\frac{M_*}{10^{10} M_\odot}\right), 0.378 \right)   \mathrm{if}\ \mathrm{SF} \\
	\mathcal{N} \left( 0.28 + 0.42 \log_{10} \left(\frac{M_*}{10^{10} M_\odot}\right) , 0.4 \right) \mathrm{if}\ \mathrm{not}\ \mathrm{SF}
	\end{cases}
\end{eqnarray}
for the disk. The medians of all four of these distributions are assumed to evolve as $r_E \propto (1+z)^{-1}$ \citep{van_der_wel_3d-hst+candels:_2014, allen_size_2017}, where equations \eqref{eq:rEb} and \eqref{eq:rEd} are evaluated at $z=1$.

\begin{figure*}
\centering
\includegraphics[width=6in]{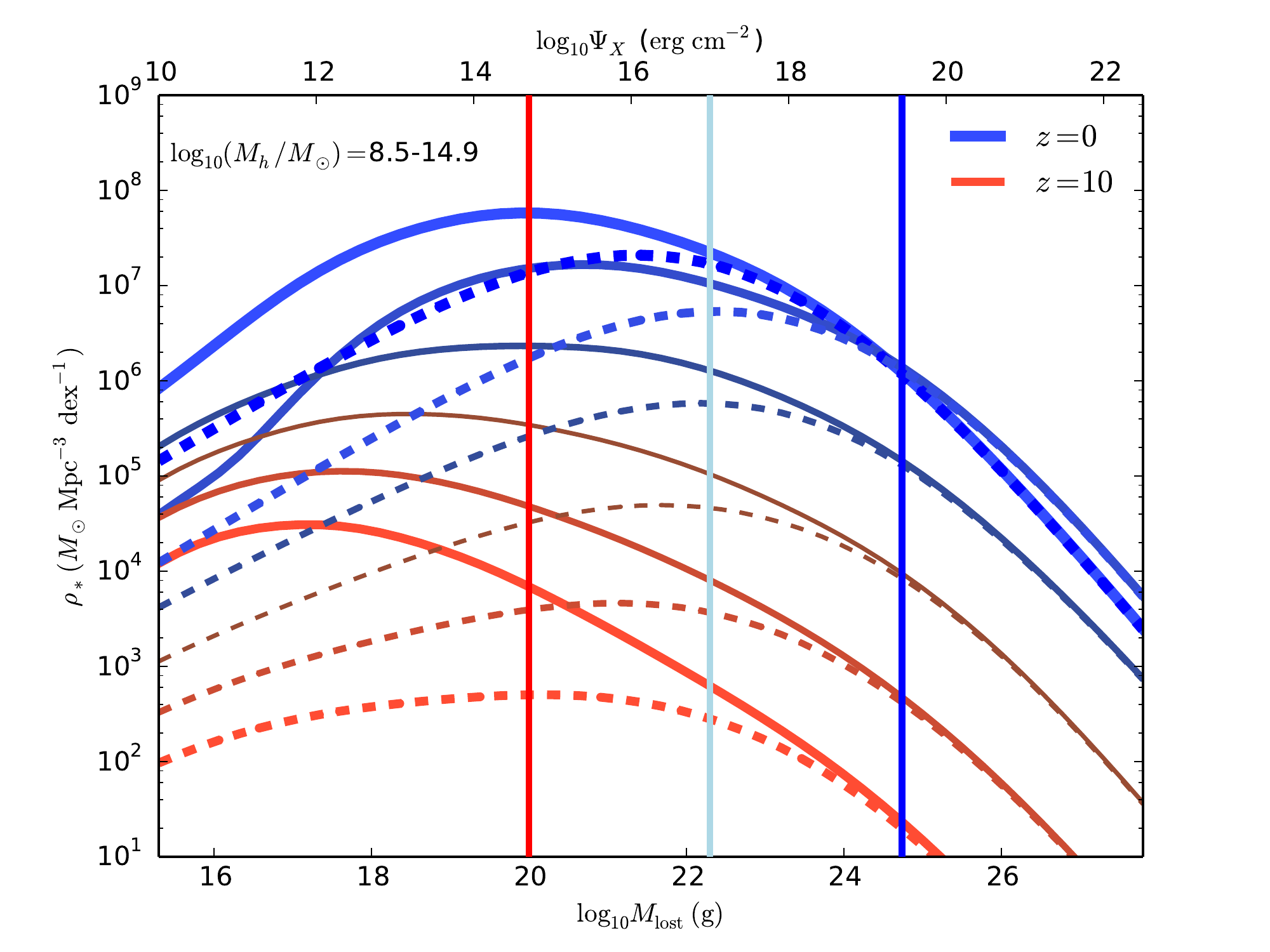}
\caption{The atmospheric mass loss distribution for all galaxies from $z=0$ to $z=10$. Here we show the result of Equation \eqref{eq:sum}, summing over all halo masses between $10^{8.5}$ and $10^{14.9} M_\odot$. Each redder line shows the result for a higher redshift, while the dashed lines show the contribution of the spheroidal components alone at that redshift. As in Figure \ref{fig:demoplot}, vertical lines show reference values of mass lost corresponding to the Martian atmosphere, Earth's atmosphere, and the mass in Earth's oceans.}
\label{fig:distr}
\end{figure*}

We carry out this procedure in a number of halo mass ranges, drawing 30,000 samples in evenly spaced intervals of $\log_{10} M_h$. The results are presented in Figure \ref{fig:distr}, which shows $d\rho_*/d\log_{10} \Psi_X$. As in Figure \ref{fig:demoplot}, we include three vertical reference lines corresponding to some relevant masses from our solar system, namely the current mass of the Martian atmosphere, the current mass of Earth's atmosphere, and the current mass of Earth's oceans. We show lines for $z=0$, 2, 4, 6, 8, and 10 with increasingly redder lines, and we also separate out the spheroidal components. As the redshift decreases, $d\rho_*/d\log_{10}\Psi_X$ generally increases at all values of $\Psi_X$ as the mass of stars in the universe increases. The contribution to the high-fluence end of the distribution is dominated by spheroids, and non-negligible parts of the distribution lie at values of $\Psi_X$ above the reference lines. Based on the drop-off in $\epsilon$ for XUV fluxes below $0.1\ \mathrm{erg}\ \mathrm{s}^{-1}\ \mathrm{cm}^{-2}$, fluences below the line for the Martian atmosphere are unlikely to correspond any longer to the values of $M_\mathrm{lost}$ on the lower $x-$axis.

\begin{figure*}
\centering
\includegraphics[width=6in, trim={0 1cm 0 0},clip]{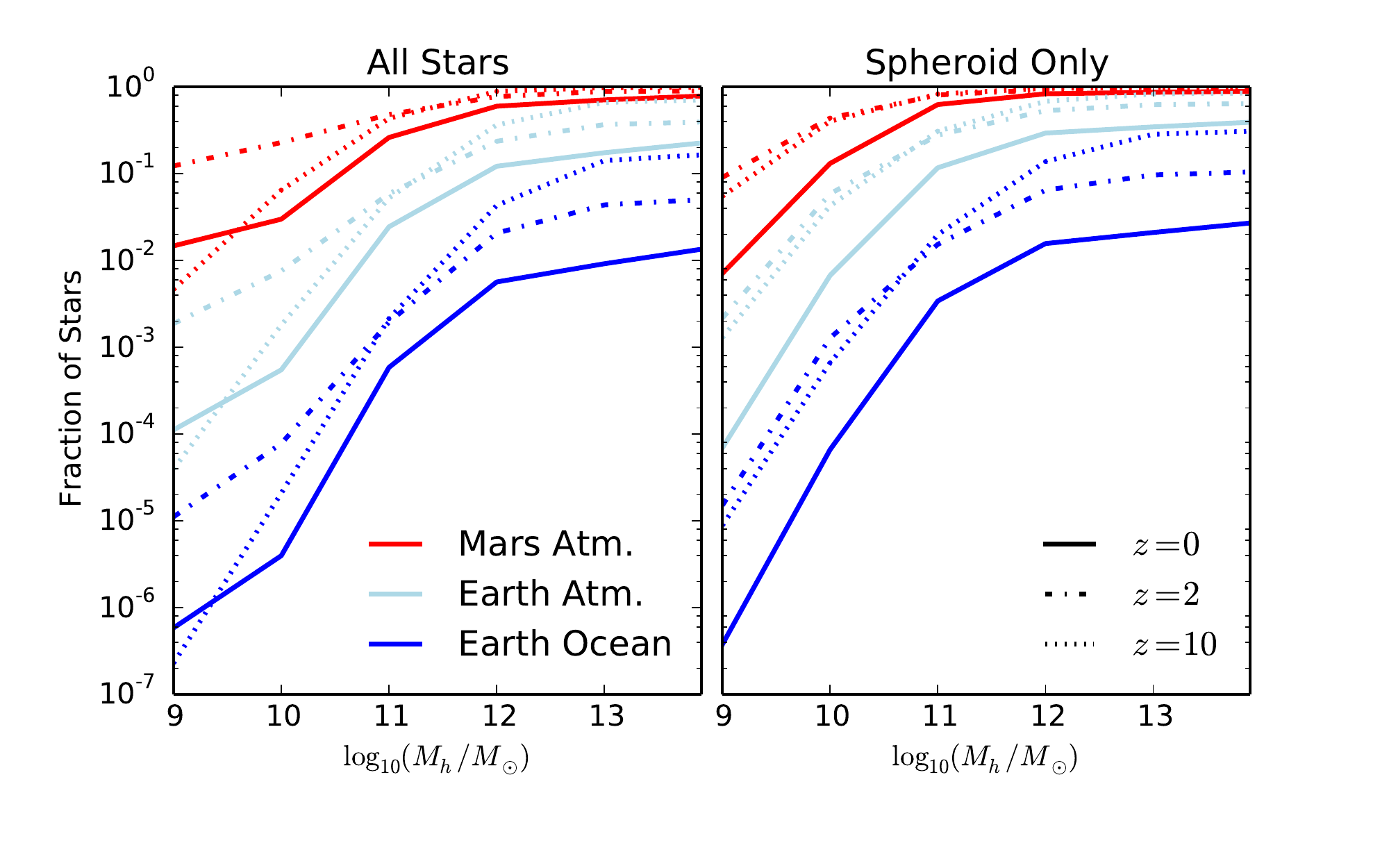}
\caption{The halo mass-dependence of the XUV fluence distribution. Each line corresponds to the fraction of stars at a given halo mass that are subject to at least enough XUV fluence to remove the given reference mass from their atmospheres. The line colors correspond to the reference masses displayed as vertical lines in Figure \ref{fig:demoplot} and \ref{fig:distr}. The line styles correspond to different redshifts. Higher mass galaxies have a larger fraction of their stars subject to any given minumum XUV fluence.}
\label{fig:vsmass}
\end{figure*}

The effect can also be broken down by mass and redshift. Figure \ref{fig:vsmass} shows the fraction of stars at a given halo mass (in a given component at a given redshift) for each of the three reference quantities of mass loss. Graphically one can imagine that these lines are where the cumulative distribution functions (CDFs) computed from Figure \ref{fig:distr} cross the vertical reference lines of that figure. Although the effect is slightly different at different redshifts, the trend from Figure \ref{fig:vsmass} is quite clear: planets in higher-mass galaxies are more likely to be subject to high XUV fluences. This is a consequence of the black hole masses in these galaxies growing super-linearly with stellar mass (since the bulge-to-total ratio increases), but the effective radii growing at a much more modest rate.

\begin{figure*}
\centering
\includegraphics[width=6in, trim={0 2cm 0 0},clip]{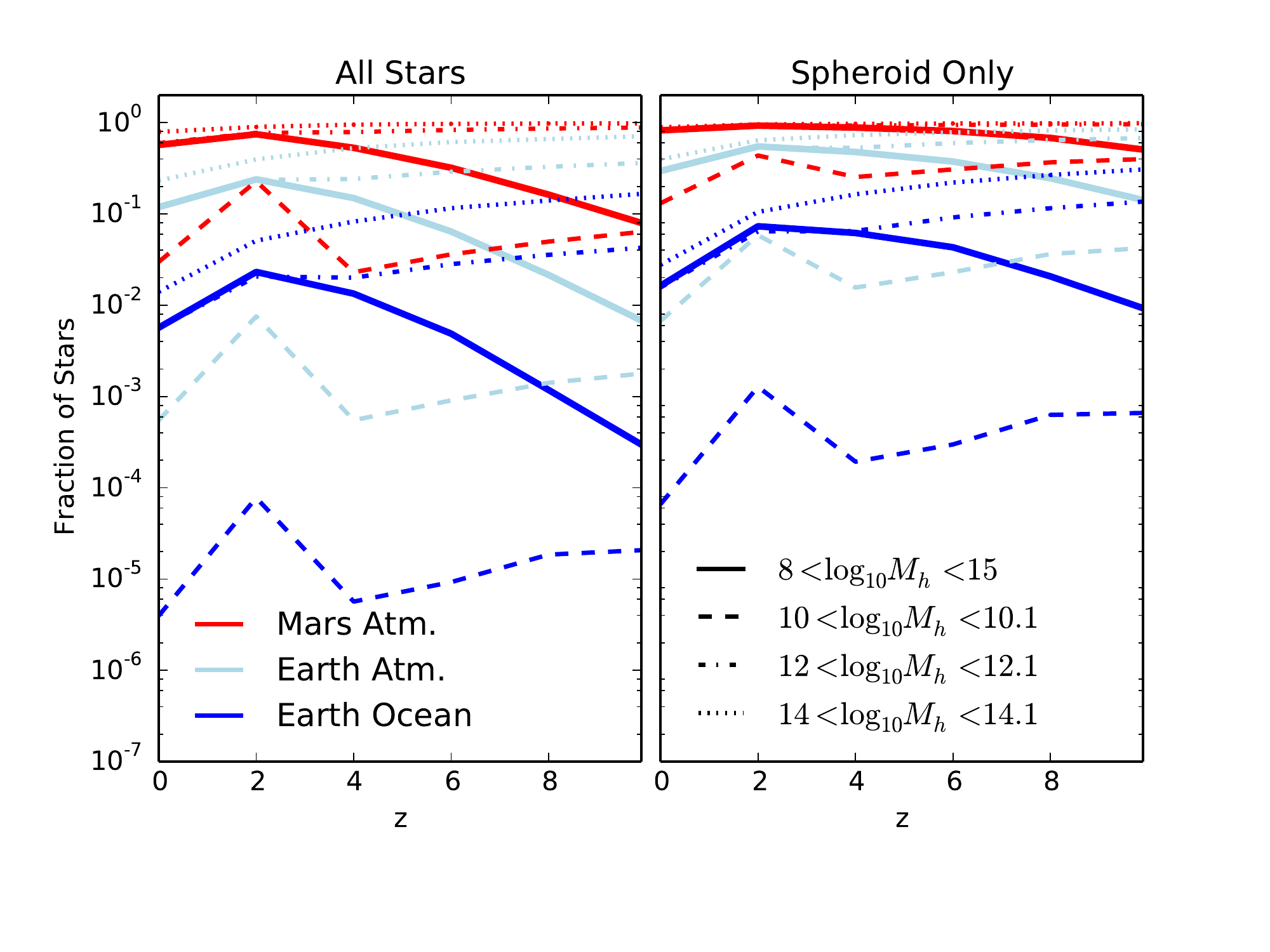}
\caption{The redshift dependence of the XUV fluence distribution. Each line shows the fraction of stars at a given redshift subject to at least enough XUV fluence to remove the reference masses (denoted by different colors) from the atmosphere of a planet. The solid lines show the fraction over all galaxies, while the dashed, dash-dotted, and dotted lines show the redshift dependence over narrow fixed ranges of halo mass. At lower redshifts, a higher proportion of the total population of stars is found in higher-mass galaxies, so while at any given mass fewer stars are subject to a given XUV fluence, over time more planets are influenced by XUV radiation from their supermassive black hole.}
\label{fig:vsz}
\end{figure*}

The redshift dependence, shown in Figure \ref{fig:vsz}, is much less dramatic. In each mass bin, shown by the non-solid lines, there tends to be a moderate decrease in the influence of XUV radiation at lower redshifts. This is the result of the increase in effective radius for galaxies at low redshift. However, over the whole mass range considered (the solid lines), the trend is reversed. Except at the very lowest redshifts, the fraction of stars in the universe subject to high XUV radiation increases with time. This apparent paradox is merely the result of the increasing characteristic mass of the halo mass function -- in other words, at low $z$, high-mass galaxies become more common, and these are the galaxies where the largest fraction of stars are subject to high XUV fluences.

\section{The influence of non-circular orbits}
\label{sec:results}

Implicit in the calculations presented in the previous section is that the stars follow roughly circular orbits, so that their current radii are representative of their typical radii. In other words, we assumed that when computing the fluence $\Psi_X = \int L_X/(4\pi r^2) dt$, the $r^2$ factor could be taken out of the integral. This is likely to be a good approximation in the disk components of galaxies, modulo the issue of stellar migration \citep[e.g.][]{sellwood_radial_2002}, but in spheroidal systems where the velocity distribution is often closer to isotropic, stars can easily traverse substantially different radii over the course of a single orbit.

To get an idea for the order of magnitude of this effect, we examine orbits in the Hernquist potential we adopted in the previous section for the density distribution of spheroidal systems. For each orbit, we compute the ratio of $\int_0^{t_0} r^{-2} dt$ to $r_0^{-2} t_0$, where $r_0$ is the particle's initial radius. The latter quantity is proportional to the fluence if the particle is on a circular orbit at radius $r_0$, while the former is the analogous quantity taking into account the fact that particle's radius changes over time. The ratio is therefore equal to $\Psi_X/\Psi_{X,\mathrm{circ}}$, the actual fluence over the fluence assuming a circular orbit.

We compute this distribution using a Monte Carlo method. In particular, we draw a large number of particles at random from the 6D phase space density $f$ of the system, and integrate them forward in time for $40 \mathrm{Myr}$. During the integrations, in addition to the particle's state, we integrate $r^{-2} dt$ to obtain the numerator of the ratio defined above. We draw particles from the phase space density in three steps. First, the particle's radius is drawn from the probability density $4\pi r^2 \rho / M_\mathrm{tot}$. Since $\rho$ has a particularly simple form for the Hernquist profile, this equation can be easily integrated and inverted to solve for $r$ as a function of the CDF, making it trivial to draw a random value of $r$ with the inverse CDF method. Once $r=r_0$ is known, we draw the magnitude of the velocity from the conditional probability density 
\begin{equation}
	p(v|r) \equiv \frac{p(v,r)}{p(r)} = \frac{4\pi r^2 4\pi v^2 f}{4\pi r^2 \rho}
\end{equation}
Once again we have made use of the spherical symmetry to integrate out the spatial angles, and similarly we assume the velocity is isotropic to integrate out the velocity angles. Note that $f$ is normalized such that $f d^3\vec{x}d^3\vec{v}$ is the mass, as opposed to the probability, contained in the differential 6D volume about a given point. For a given $r$ and $v$, this expression can be evaluated analytically using the expression from \citet{hernquist_analytical_1990} for $f$ as a function of the energy per unit mass $E = (1/2) v^2 + \phi(r)$ where $\phi(r)=GM/(r+a)$ is the gravitational potential. Note that $E$ is strictly a function of $r$ and $v$. We sample from $p(v|r)$ using rejection sampling from a uniform distribution, making use of the fact that $p(v|r)$ is a broad distribution, and there is a finite range of allowed velocities, namely $0<v<\sqrt{2 G M_\mathrm{tot}/(r+a)}$, for which a star is bound. Finally, the star can be assigned a velocity vector $\vec{v}$ by drawing a random point from the surface of the unit sphere, again by the assumption of isotropy.

Once the star's initial radius and velocity vector are chosen, we immediately know the star's specific energy (as in the previous paragraph) and initial specific angular momentum $L= r(v_x+v_y)$ where, because of the spherical symmetry, without loss of generality we have assumed that the star lies on the $z$-axis. On timescales similar to or shorter than the dynamical time, as is the case for our integration, $E$ and $L$ are constants. We can therefore compute $r$ as a function of time by integrating
\begin{equation}
	\frac{dr}{dt} = v_r = \pm \sqrt{2(E-\phi(r)) - \frac{L^2}{r^2}}.
\end{equation}
 The two solutions correspond to the particle moving inward and outward along its orbit. In order to choose the correct root, we flag the initial sign of $v_r$ as determined by the initial draw of the velocity, and we change the sign when the argument of the square root becomes negative. In this unphysical regime, we evaluate the square root using the absolute value of the argument, and the particle quickly re-enters the physically allowed region. We verify that the distribution of $r$ for a large sample of these particles agrees with the analytic distribution both in the initial conditions and after the $40$ Myr evolution.

\begin{figure}
\centering
\includegraphics[width=3.7in]{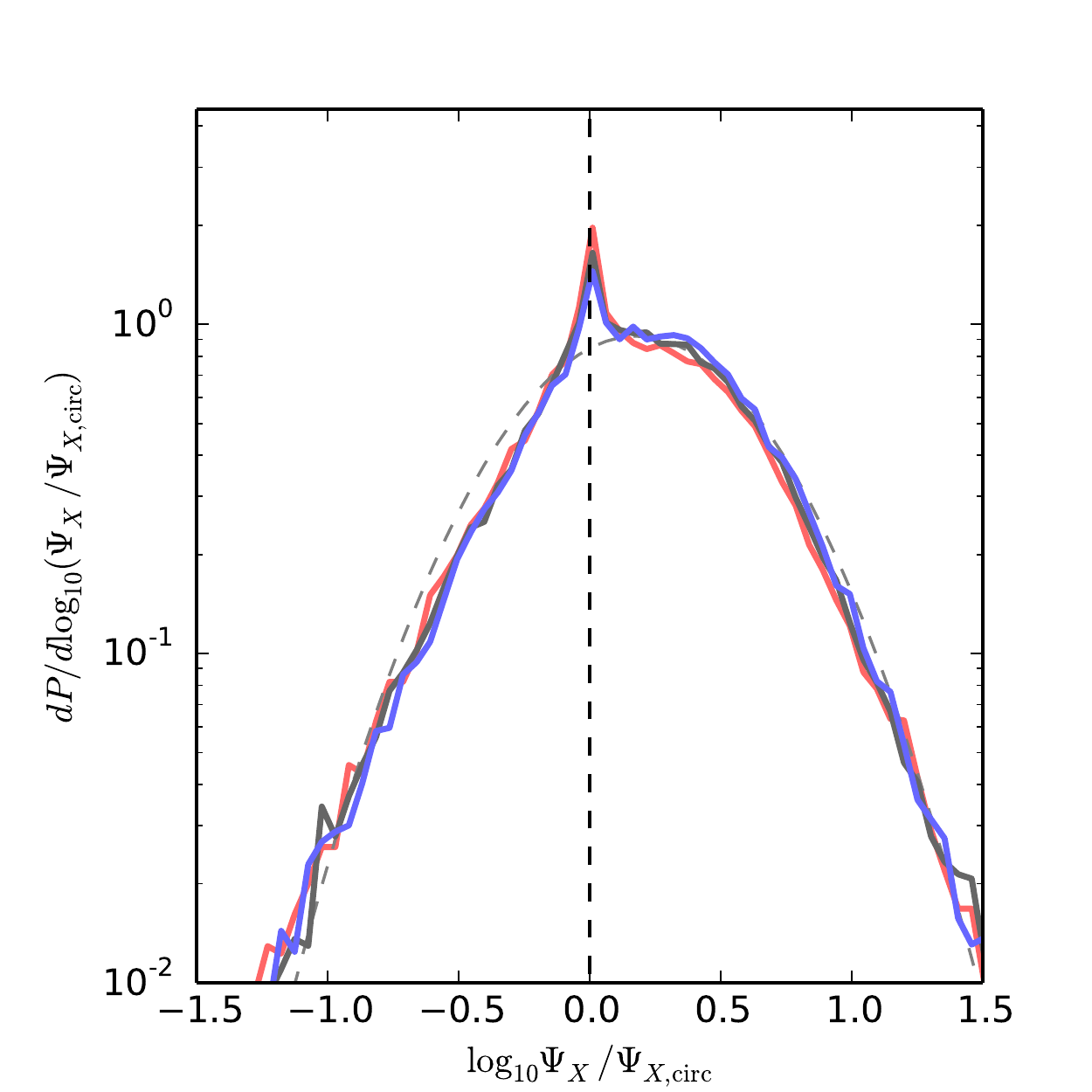}
\caption{The probability distribution of fractional fluence enhancements relative to the case of circular orbits. The vertical dashed line shows $\Psi_X/\Psi_{X,\mathrm{circ}}=1$, corresponding to no enhancement of the fluence relative to the case of circular orbits. Each color from red to blue shows a different bulge mass from $10^8 M_\odot$ to $10^{12} M_\odot$ in steps of 2 dex. There is a narrow peak containing about $5\%$ of the mass at $\Psi_X/\Psi_{X,\mathrm{circ}}=1$ for stars that happen to be near circular orbits. Excluding this peak, the remainder of the distribution is approximately fit by a log-normal, shown as a gray dashed line. Fluences are likely to be enhanced by a factor of $50\%$ with deviations up to $\Psi_X/\Psi_{X,\mathrm{circ}} = 4$ and down to a fluence decrease of a factor of two. There is only subtle dependence on bulge mass over the four orders of magnitude included here.}
\label{fig:noncirc}
\end{figure}

The result of this calculation is shown in Figure \ref{fig:noncirc} for masses between $10^8 M_\odot$ and $10^{12} M_\odot$, and $a$'s set according to the median of the distribution for quenched galaxies in equation \eqref{eq:rEb}. Although there is a narrow peak at $\Psi_X/\Psi_{X,\mathrm{circ}} = 1 $ containing about 5\% of the stars, the full distribution is much broader, with a distribution reasonably fit by a log-normal with median $\log_{10} \Psi_X/\Psi_{X,\mathrm{circ}} \approx 0.18$ and width $0.43$ dex. The distribution is also roughly independent of mass. Qualitatively, the distribution of $\Psi_X$ itself is so broad that the additional width introduced by the effect of non-radial orbits is not likely to be important.

\section{Implications for Habitability}
\label{sec:discussion}

The habitability of any given planet is complex and time-dependent. In some sense, one must understand the planet's climate over Gyr timescales where there are many more unknowns than in terrestrial climate science -- the planet's mass, spin, inclination, period, surface configuration, atmospheric composition and pressure, and so forth \citep[see section 4.1 of][and references therein]{shields_habitability_2016}. In this section we shall nonetheless comment on a few aspects of habitability implied by our results.

\subsection{Galactic Habitable Zones}
For many readers, our results may be reminiscent of the concept of the galactic habitable zone \citep{ gonzalez_galactic_2001, lineweaver_galactic_2004, forgan_evaluating_2017, spitoni_galactic_2017}. Just as each star has a circumstellar habitable zone, a set of planetary orbits that are likely to admit stable climates with liquid water at the surface \citep{kopparapu_habitable_2013}, it has been suggested that each galaxy has a particular range of radii where life is more likely to develop. In most formulations, it is assumed that the metallicity of the gas from which a planetary system forms must be in a certain range near the solar value. This supposition is based on the fact that terrestrial planets form from heavy elements, so planets are more likely to form when the metallicity is higher. If the metallicity is too high, planet formation may make gas giants more common, leading to the ejection of many terrestrial planets \citep[e.g.][]{barclay_demographics_2017}. Suppositions about habitability based on metallicity are quite speculative, since observationally there is no indication of any dependence of terrestrial planet frequency on metallicity \citep{buchhave_abundance_2012, schlaufman_continuum_2015}. There is a well-known correlation between metallicity and gas giant frequency \citep{fischer_planet-metallicity_2005}, but the consequences for habitability are unclear.

Another key component of galactic habitable zones is the proximity of the planet to supernovae and gamma-ray bursts. Most stellar explosions occur in proportion to the star formation rate averaged over at least 40 Myr timescales, so the centers of galaxies, particularly near $z=2$ are supposed to be particularly dangerous \citep{lineweaver_galactic_2004} though the absolute normalization of the effect, i.e. the exact probability that a supernova at a given distance will render a planet uninhabitable, is quite uncertain. A useful starting point is the computation by \citet{gehrels_ozone_2003}, which examines the effects on Earth's atmosphere by running a global circulation model coupled to a photochemical network. The primary effect of a supernova according to this result is the production of odd nitrogen molecules (namely a single nitrogen paired with some number of oxygen atoms), which catalyze the destruction of ozone. The magnitude of the effect is such that a supernova at $\sim 8\ \mathrm{pc}$ would cause order unity destruction of the Earth's ozone layer over a multi-year period. Although this would undoubtedly cause damage to Earth's current biosphere, it is less clear that it would render the Earth uninhabitable, much less a generic terrestrial planet, whose atmospheric chemistry and thickness may be substantially different.

The mechanism we discuss in this work, namely the loss of substantial atmospheric mass due to winds driven by XUV radiation from AGN, is arguably a much stronger and more robust effect than the ozone destruction by supernovae typically considered. However, it is unclear whether the levels of mass loss we find in this paper are in fact destructive for life. A useful comparison is Proxima Centauri b, whose host star is prone to X-ray flares, and is therefore likely to have experienced a large fluence of high-energy radiation over the course of its life. In spite of this substantial mass loss, there are many scenarios \citep{ribas_habitability_2016, barnes_habitability_2016} where the planet retains substantial volatiles. On the other hand, it has recently been shown \citep{jakosky_mars_2017} that mass loss from the Martian atmosphere (through a different mechanism) was likely responsible for a dramatic change in character of the planet, from temperate with substantial liquid water on the surface, to the current cold dusty planet where nearly all of the water near the surface is frozen. It is likely that two different planets subject to the same atmospheric mass loss will experience dramatically different climate and habitability effects, ranging from negligible to dramatic. It is also possible that atmospheric mass loss may increase the habitability of a planet, if the planet initially forms with a substantial hydrogen atmosphere \citep{luger_habitable_2015}.

\subsection{Implications for Earth}

As shown in Table \ref{tab:sources}, the high-energy flux at Earth's current distance from the Milky Way's central supermassive black hole is small, even if Sgr A$^*$ were emitting an unobscured Eddington luminosity. It follows that even though there are multiple lines of evidence pointing to recent accretion events \citep{su_giant_2010, bland-hawthorn_fossil_2013, churazov_not_2017}, planets in the solar system itself have probably never been subject to this particular mechanism of mass loss. There is at least one caveat to this conclusion, however. While stars in disk galaxies typically follow circular orbits, they may be perturbed when they pass through resonances with spiral arms or bars \citep{sellwood_radial_2002}. Simulations show that the metallicity distribution function of the Milky Way has properties (in particular skewness) consistent with a substantial fraction of stars in the outer disk having experienced migration \citep{loebman_imprints_2016}. In fact, the high measured metallicity of Alpha Centauri has led \citet{barnes_habitability_2016} to suggest that the system has migrated outwards by at least 3.5 kpc. The Sun may have experienced such a process, but it is also quite consistent with having been born near its current galactocentric radius \citep{allende_prieto_s4n:_2004}. Given the mass of Sgr A$^*$, and using $\Psi_X \sim 10^{17} \mathrm{erg}\ \mathrm{cm}^{-2}$ in Equation \eqref{eq:rcrit}, planets in the Milky Way are only likely to be affected if their stars orbit within $\sim 0.1 $ kpc of the galactic center. It follows that even planets in the Alpha Centauri system, where there is evidence for substantial past radial migration, are unlikely to have been affected by mass loss due to XUV irradiation by Sgr A$^*$.

Another possibility arises during the course of a merger between two galaxies. Each one initially contains a black hole at its center. Over time, the black holes will be driven to the center of the post-merger galaxy by dynamical friction \citep{begelman_massive_1980, kelley_massive_2017} on a timescale comparable to the dynamical time. During this phase, if the black holes are actively accreting, they could essentially extend the reach of their XUV radiation far beyond the center of the galaxy. There are numerous examples of pairs of AGN on kpc scales \citep{fu_nature_2012, mcgurk_spatially_2015, comerford_merger-driven_2015}. If the black holes are no more likely than usual to be active as they settle to the center of the galaxy, then the small quasar duty cycle would suggest that this is a minor effect. However, mergers are exactly when the black holes may be most likely to be active. Even so, the effect of these off-center AGN is not likely to be larger than the central AGN simply because the stellar density at the center of the galaxy is larger than in its outskirts.

Another aspect of mergers is the rearrangement of stellar orbits. Although the Milky Way likely has not had a substantial merger in the past 10 Gyr \citep{stewart_merger_2008}, it is destined to merge with Andromeda in a few Gyr \citep{cox_collision_2008, sohn_m31_2012, van_der_marel_m31_2012}. Simulations with orbital parameters consistent with constraints on the relative motion of the Milky Way and Andromeda predict that the merger remnant will be a large elliptical galaxy. The simulations also predict the distribution of final orbits for stars at similar present-day galactocentric radius to the sun. These stars are predicted to have substantially larger semimajor axes than $\sim 8$ kpc \citep{van_der_marel_m31_2012}, suggesting once again that solar system planets are unlikely to ever face enough XUV radiation from a supermassive black hole to harm their atmospheres.

\section{Conclusion}

In this paper we have brought together three different astrophysical subjects: the accretion history of supermassive black holes, the spatial distribution of stars in galaxies, and the erosion of planetary atmospheres. We assumed that most galaxies in the universe have a supermassive black hole which has grown largely through the direct accretion of gas, largely over a small number of instances where the accretion luminosity reaches the Eddington limit for a short time, about 40 Myr. Next, we followed observational relationships between stellar and halo mass, bulge to total ratio, the spatial sizes of the bulge and disk, and black hole mass. From these two sets of assumptions, we can compute the PDF of XUV fluences to which random stars in the universe are subject by adding up the analytically-determined PDFs associated with individual stellar components. The scatter in each observational relation is included by a Monte Carlo method, in which the sum is carried out over many draws from each distribution.

The atmospheric mass loss PDF can be computed at a variety of epochs, and the contribution to the PDF from bulges vs. disks, and from different ranges of halo mass can be tracked as well. We have found that the largest fluences are contributed by the highest-mass halos, which have the largest black holes, but not dramatically larger spatial sizes. The PDF changes relatively little across cosmic time because, although galaxies at a given mass are growing in spatial size, the increasing contribution of higher halo masses as the characteristic halo mass increases largely cancels out the effect. In the end the XUV fluence from supermassive black holes is appreciably higher in the spheroidal components of galaxies, and depends strongly on galaxy mass, but only weakly on redshift.

The XUV fluence can be directly related to the mass of atmosphere lost by a planet of a given density over a range of about 4 orders of magnitude in fluence. Using this translation, we find that a substantial fraction of planets in the universe may have lost potentially non-negligible quantities of atmosphere (see Figure \ref{fig:vsz}). In particular, about 50\% of all planets in the universe may lose the equivalent of a Martian atmosphere, 10\% may lose an Earth's atmosphere, and 0.2\% may lose the mass of Earth's oceans. Although ultimately the effect on habitability is uncertain, we note that the damage inflicted to planetary atmospheres by XUV irradiation from black holes may be substantially greater than what some past works have assumed would render the Earth uninhabitable.

\section*{Acknowledgements}

The authors would like to thank Lars Hernquist, Carl Rodriguez, James Guillochon, James Owen, Aaron Romanowsky, and Rory Barnes for helpful discussions. This work has benefitted from numerous freely available resources, including the arXiv, NASA ADS, numpy, scipy, and matplotlib; we thank those responsible for making these resources available. JCF is supported by an ITC Fellowship.

%\bibliography{libJul3}
\bibliography{/Users/jforbes/updatingzotlib}

\appendix

\section{Habitable zones around compact objects}

The first exoplanets ever discovered were found orbiting a millisecond pulsar \citep{wolszczan_planetary_1992}, and the case for searching for planetary transits around X-ray binaries has recently been made by \citet{imara_searching_2017}. The influence of XUV radiation of such sources on nearby planets is therefore an interesting question. In particular, how much damage is inflicted on the atmosphere of a planet in the habitable zone of an object emitting high energy radiation?

Since the location of the habitable zone is, to a first approximation, just set by the bolometric flux the planet receives, the corresponding XUV flux is just the fraction of the bolometric luminosity emitted in high-energy radiation, $\eta_X$, times the flux corresponding to habitability. For reference, the bolometric flux at Earth is $F_E = 1.4 \times 10^7 \mathrm{erg}\ \mathrm{cm}^{-2}\ \mathrm{s}^{-1}$. For $\eta_X \ga 10^{-3}$, we are well within the regime where the mass loss scales as $F_X^{1/2}$ instead of directly with $F_X$. Following the results of \citet{bolmont_water_2017}, we will reduce $\epsilon$ by $\sqrt{F_X/10^{3}\ \mathrm{erg}/\mathrm{cm}^2/\mathrm{s}}$, in which case the mass lost over a period of 1 Gyr is
\begin{equation}
	M_\mathrm{lost} = 7.7\times 10^{25}\ \mathrm{g} \left(  \frac{\eta_X}{0.1} \frac{F_\mathrm{bol}}{F_E} \right)^{1/2} \frac{t_0}{1\ \mathrm{Gyr}}
\end{equation}
It follows that for planets around such objects, water may be liquid on their surfaces, but it is unlikely to remain there for very long.

\clearpage

\end{document}